\documentstyle[latexsym,amssymb,epsfig,aps,floats,eqsecnum,amsfonts,twoside]{revtex}
\newtheorem{THEOREM}{Theorem}
                        {\end{THEOREM}}
\newtheorem{PROPOSITION}[THEOREM]{Proposition}
\newenvironment{proposition}{\begin{PROPOSITION} \hspace{-.85em}
{\bf :} }%
                            {\end{PROPOSITION}}
                      {}

\newcommand{\pro}{\begin{proposition}}
\newcommand{\epro}{\end{proposition}}
\newcommand{\prf}{\noindent{\bf Proof:} }
\newcommand{\qed}{$\blacksquare$}

\def\br{\begin{eqnarray}}
\def\er{\end{eqnarray}}
\def\brn{\begin{eqnarray*}}
\def\ern{\end{eqnarray*}}
\def\er{\end{eqnarray}}
\def\beq{\begin{equation}}
\def\eeq{\end{equation}}

\def\vt{\vartheta}

\def\L{{\cal{L}}}

\def\d{\delta}

\def\th{\theta}
\def\s{\sigma}

\def\t\s{\tilde{s}}

\def\L{{\mathcal{L}}}
\def\E{{\mathcal{E}}}
\def\T{{\mathcal{T}}}

\def\C{{\mathcal{C}}}
\def\F{{\mathcal{F}}}
\def\M{{\mathcal{M}}}

\def\X{{\mathcal{X}}}
\title{ Weak field reduction in teleparallel coframe gravity. Vacuum case.}
\author{
\thanks {\quad   email itin@math.huji.ac.il}
\small {\rm Yakov Itin}\\
\small {\rm Institute of Mathematics}
\small {\rm Hebrew University of Jerusalem}\\
\small {\rm Givat Ram, Jerusalem 91904, Israel}\\
}
\begin{document}

\pagestyle{myheadings}
\markboth{Y. Itin \qquad}{Weak field reduction in coframe gravity}

\date{\today}
\maketitle

\begin{abstract}

The  teleparallel coframe gravity may be viewed as a
generalization of the standard GR. 
A coframe (a field of four independent 1-forms) is considered, 
in this approach,  to be a basic dynamical variable. 
The metric tensor is treated as a secondary structure. 
The general Lagrangian, quadratic in the first order derivatives of the 
coframe field is not unique. 
It involves three dimensionless free parameters. 
We consider a weak field approximation of the general coframe teleparallel
model.
In the linear approximation, the field variable, the coframe, is covariantly
reduced to the superposition of the symmetric and antisymmetric field.
We require this reduction to be preserved  
on the levels of the Lagrangian, of the field equations and of the conserved currents. 
This occurs if
and only if the pure Yang - Mills type term is removed from the Lagrangian.
The  absence of this term is known to be necessary and sufficient for the 
existence  of the viable (Schwarzschild) spherical-symmetric solution. 
Moreover, the same condition guarantees the absence of ghosts and tachyons in particle 
content of the theory. The condition above is shown recently to be necessary for 
a well defined Hamiltonian formulation of the model. 
Here we derive the same condition in the Lagrangian formulation by means of the 
weak field reduction.  

\end{abstract}
\section{Introduction}         \label{sec:intro}      
Einstein's general relativity (GR) is very successful in describing the
long distance (macroscopic) gravity phenomena.
This theory, however, encounters serious difficulties on microscopic distances. 
So far essential problems appear in all attempts to quantize the standard GR
(for recent review, see, e.g., \cite{Carlip:2001wq}).
Also, the Lagrangian structure of GR differs, in principle, from the ordinary
microscopic gauge theories.
In particular, a covariant conserved energy-momentum tensor for the gravitational field
cannot be constructed in the framework of GR.
Consequently, the study of alternative models of gravity is justified
from the physical as well as from the  mathematical point of view.
Even in the case when GR is unique true theory of gravity,
consideration of close alternative models can shed light on the properties of GR itself.

Among various alternative constructions, the Poincar{\'e} gauge theory of
gravity, see Refs. \cite{Hehl:kj} --- \cite{Yo:2001sy},
is of a special interest.
This theory proposes a natural bridge between gauge and geometrical theories. 
Moreover, it has a straightforward generalization to the
metric-affine theory of gravity \cite{Hehl:ue}, which involves a wide spectra 
of spacetime geometries. 
However, it was elucidated recently that even the restriction of the Poincar{\'e}
gauge theory to the teleparallel model provides a reasonable alternative to GR, 
see e.g. \cite{Itin:2001bp}. 
\subsection{Coframe (teleparallel) gravity --- basic facts and notations}
We start with a brief account of the coframe (teleparallel) model of gravity and 
establish the notations used in this paper. 
Details, different approaches  and additional references can be found in
\cite{Hayashi} -- \cite{Itin:1999wi}.

\noindent Let a $4D$ differential manifold $\M$ be endowed with two smooth fields:
a frame field $e_a$ and a coframe field $\vt^a$.
In a local coordinate chart,
\begin{equation}\label{tg.0}
e_a={e_a}^\mu(x) \,\, {\partial}/{\partial x^\mu}\,,\qquad
\vt^a={\vt^a}_\mu(x) \,\, dx^\mu\,,\qquad a,\mu=0,1,2,3\,.
\end{equation}
These fields allow to compare two vectors (more generally, two tensors)
attached to different points of the manifold. It is referred to as the
{\it teleparallel structure} on $\M$.
The two basic fields are  assumed to fulfill the dual relation: $e_a\rfloor \vt^b=\d_a^b$.
We denote by  $\rfloor$ the interior product operator 
$\X\times \Lambda^p\to  \Lambda^{p-1}$
 that, for an arbitrary vector field $X\in\X$ and  a $p$-form field $w\in\Lambda^p$,
  $X\rfloor w:=w(X,\cdots)$.
So only one of the fields, $e_a$ or $\vt^a$, is 
independent. Thus, two alternative (but, principle, equivalent)
representations of the teleparallel geometry are possible.

\noindent The {\it frame representation} is based on a complex
$\{\M,e_a\}$ and applies the tensorial calculus as the main
mathematical tool similar to  the Einstein tensorial representation of GR.

\noindent The  {\it coframe  representation}, which  deals with  a complex $\{\M,\vt^a\}$, 
applies the exterior form technique.
In present paper, we use this approach and call it the {\it coframe
gravity}, in contrast to the metric gravity of GR.

In a wider context, the coframe field appears  as one of the basic dynamical
variables in the Poincar{\'e} gauge gravity  and in the metric-affine gravity.
To extract the pure coframe sector, in these theories, one has
to require vanishing of the  curvature. 
Here, we  treat the coframe field as a  self-consistent
dynamical variable with its own covariant operators: wedge product, Hodge map
and exterior derivative.
These  two approaches (one with a trivial connection and the other
without  explicit exhibition of a connection) are principally equivalent.

The indices in (\ref{tg.0}) are basically different. The Greek indices refer to the
coordinate space and describe the behavior of tensors under the group of
diffeomorphisms of the manifold $\M$.
The Roman indices denote different 1-forms of the coframe.
The corresponding group of transformations, $SO(1,3)$, comes together with its
natural invariant $\eta_{ab}=diag(1,-1,-1,-1)$. 

The metric tensor on $\M$ is expressed  via the coframe as
\begin{equation}\label{tg.1}
g=\eta_{ab}\vt^a\otimes\vt^b\,,
\end{equation}
i.e., the coframe is postulated to be pseudo-orthonormal.
The coframe field   and all the objects constructed from it are
 assumed to be global (rigid) covariant.
In other words, all the  constructions are required to be covariant under
the global transformations $\vt^a\to {A^a}_b\vt^b$ with a constant matrix 
 $ {A^a}_b\in SO(1,3)$.
The  metric tensor (\ref{tg.1}) is invariant under
a wider group of transformation:
local (pointwise) transformations of the coframe with ${A^a}_b={A^a}_b(x)$.

Consider a Lagrangian density, which is  (i) diffeomorphism invariant,
(ii) invariant under global $SO(1,3)$ transformations of the coframe and
(iii) quadratic in the exterior derivatives of the coframe. 
The most general  Lagrangian of this form  is a linear combination
 \cite{Itin:2001bp}, \cite{Muench:1998ay}, 
\begin{equation}\label{tg.5}
\L=\frac 12 \sum_{i=1}^3 \rho_{i} \; {}^{(i)}\L\,,
\end{equation}
where $\rho_1,\rho_2,\rho_3$ are free dimensionless parameters.
The  linear independent 4-forms appearing here are expressed via the 
coframe {\it field strength}, $\C^a:=d\vt^a$.
\br\label{tg.2}
{}^{(1)}\L &=&\C^a \wedge *\C_a\,,\\
\label{tg.3}
{}^{(2)}\L &=&\left(\C_a \wedge \vt^a\right)  \wedge*\left(\C_b\wedge\vt^b\right)\,, \\
\label{tg.4}
{}^{(3)}\L &=& \left( \C_a \wedge\vt^b\right)  \wedge *\left(\C_b \wedge \vt^a \right)\,,
\er
The Hodge dual operator $*$ is defined by the pseudo-orthonormal coframe $\vt^a$
or, equivalently, by the metric (\ref{tg.1}). 
One may try to include in the Lagrangian some invariant expressions of
the  second order   (similarly to the Hilbert-Einstein Lagrangian).
Such terms, however, are reduced to total derivatives  and
do not affect the field equations and the Noether conserved currents.
So (\ref{tg.5}) is the most general Lagrangian that generates the field equations 
of the second order. 

Let us introduce the notion of the {\it  field strength}
\begin{equation}\label{tg.6}
\F^a:={}^{(1)}\F^a+{}^{(2)}\F^a+{}^{(3)}\F^a\,,
\end{equation}
with
\br\label{tg.7}
{}^{(1)}\F^a&:=&(\rho_1+\rho_3)\C^a\,,\\
\label{tg.8}
{}^{(2)}\F^a&:=&\rho_2e^a\rfloor(\vt^m\wedge \C_m)\,,\\
\label{tg.9}
{}^{(3)}\F^a&:=&-\rho_3\vt^a\wedge(e_m\rfloor \C^m)\,.
\er
Such separation of the strength $\F^a$ involves two scalar-valued forms 
 $\vt^m\wedge \C_m$ and $e_m\rfloor \C^m$. So some calculations are simplified. 
For irreducible decomposition of  $\F^a$, see  \cite{Hehl:ue} and \cite{Itin:2001bp}. 
 
In the notation (\ref{tg.7} --- \ref{tg.9}) , the coframe Lagrangian (\ref{tg.5}) takes a form similar 
to the Maxwell Lagrangian, 
\begin{equation}\label{tg.10}
\L=\frac 12\, \C^a\wedge *\F_a\,.
\end{equation}
The free variation of  (\ref{tg.10}) relative to the coframe $\vt^a$
has to take into account also the variation of the Hodge dual operator,
which implicitly depends on the coframe. It yields the field equation of the form 
\cite{Itin:2001bp}
\begin{equation}\label{tg.11}
d*\F^a=\T^a\,,
\end{equation}
where  the 3-form $\T^a$ is the energy-momentum current of the coframe field
\begin{equation}\label{tg.12}
\T_a=(e_a\rfloor \C_m)\wedge *\F^m -e_a\rfloor \L\,.
\end{equation}
The conservation law for this 3-form: $d\T_a=0$  is a straightforward
consequence of (\ref{tg.11}).

\subsection{Viable models --- a problem of physical motivation}
A general quadratic coframe model, which is global  
$SO(1,3)$ invariant, involves three parameters:
\begin{equation}\label{add1-1}
\rho_1\,,\quad\rho_2\,, \quad\rho_3\quad  \!\!\!-\!\!\!-\!\!\!-
\quad  {\textrm { free}}\,.
\end{equation}
The ordinary GR is extracted from this  family  by
requiring of the {\it local} $SO(1,3)$ invariance, which is realized by 
the following restrictions of the parameters:
\begin{equation}\label{add1-2}
\rho_1=0\,,\quad 2\rho_2+ \rho_3=0 \,.
\end{equation}
The analysis of exact solutions \cite{Itin:1999wi} to the field equation (\ref{tg.11})
shows that the Schwarzschild solution appears even for a wider 
set of parameters (viable set): 
\begin{equation}\label{add1-3}
\rho_1=0\,,\quad\rho_2\,,\quad\rho_3\quad  -\!\!\!-\!\!\!-\!\!\!-
 \quad {\textrm{free}}\,.
\end{equation}
Moreover, for $\rho_1\ne 0$, spherical-symmetric static solutions to (\ref{tg.11}) 
do not have the Newtonian behavior at infinity \cite{Itin:1999wi}. 
 
\noindent So a problem arises: {\it Which physical motivated requirement
extracts the viable set of parameters?}

The quantum-theory solution to this problem is known for a long time. 
In \cite{VanNieuwenhuizen:fi} --- \cite{Shapiro:2001rz}  
 it was shown that the requirement (\ref{add1-3}) 
is necessary and sufficient for absence of ghosts and tachyons in particle 
content of the theory. Another motivation for (\ref{add1-3}) 
comes from the requirement that the theory has to have a well 
defined Hamiltonian formulation (\cite{Sousa:2003fv}). 

In this paper we look for a motivation of (\ref{add1-3}) on a classical Lagrangian level. 
We deal with  linear approximation of the general coframe
model. The coframe variable can be treated, in this approximation,
as a regular $4\times 4$ matrix. Consequently, it reduced to a composition 
of two independent variables: the symmetric and the antisymmetric fields. 

Our main result is as follows: Only for  (\ref{add1-3}), the coframe 
model is reduced to two independent models,
every one with its own Lagrangian, field equation, and conserved
current. In other words, the viable model is exactly this one that approaches the
{\it free-field limit}, i.e., any interaction between the approximately
independent fields appears only in higher orders. 

 Linear approximation of coframe models was usual applied  for study the deviation 
 of teleparallel gravity from the standard GR, and for comparison with the 
 observation data, 
 see \cite{Schweizer:1979up}, \cite{Nitsch:1979qn}, \cite{Sezgin:zf}, 
 \cite{Kuhfuss:1986rb}. In our approach the reduction of the lower order terms is 
 used as a theoretical device. We show that this condition is enough to distinguish 
 the set of viable models. The relation between these two approaches requires 
 a further consideration. 

\section{Weak field reduction} 
\subsection{Linear approximations}      
To study the approximate solutions to (\ref{tg.11}), we start with a
trivial exact solution, a {\it holonomic coframe}, for which, 
\begin{equation}\label{la.1}
d{\vt}^a=0\,.
\end{equation}
Consequently, $\F^a=\C^a=0$, so both sides of Eq. (\ref{tg.11})  vanish.
By Poincar{\'e}'s lemma, the solution of (\ref{la.1}) can be locally expressed as
${\vt}^a=d\tilde{x}^a(x)$, where  $\tilde{x}^a(x)$ is a set of four
smooth functions defined in a some neighborhood $U$ of a point $x\in\M$.
The functions $\tilde{x}^a(x)$,
being treated as  the components of a coordinate map
$\tilde{x}^a:U \to{\mathbb R}^4$, generate a local coordinate system on $U$.
The metric tensor (\ref{tg.1}) reduces, in this
coordinate chart,  to the  flat Minkowskian metric 
$g=\eta_{ab}d\tilde{x}^a\otimes d\tilde{x}^b$.
Thus the holonomic coframe plays, in the teleparallel background, the same role as the
Minkowskian metric in the (pseudo-)Riemannian geometry.
Moreover, a manifold endowed with a (pseudo-)orthonormal holonomic
coframe is  flat. 
The weak perturbations of the basic solution $\vt^a=dx^a$ are  
\begin{equation}\label{la.2}
\vt^a=dx^a+h^a=(\d^a_b+{h^a}_b)\,dx^b\,.
\end{equation}
``Weak'' means:
\begin{equation}\label{la.3}
||{h^a}_b||=\epsilon=o(1)\,, \quad 
||{h^a}_{b,c}||=O(\epsilon)\,,\quad ||{h^a}_{b,c,d}||)=O(\epsilon) \,,
\end{equation}
where $||\cdots||$ denotes the maximal tensor norm.
We accept that  the coframe $\vt^a$  and the holonomic coframe $dx^a$
have the same physical dimension of $[length]$.
Thus,  the components of the matrix ${h^a}_b$ and the parameter  $\epsilon$
are dimensionless.  Consequently,
the approximation conditions   (\ref{la.3}) are  invariant
under rescaling of the coordinates.

In this paper we will take into account only 
the first order approximation in the perturbations ${h^a}_b$ and
in their derivatives (i.e., in the parameter $\epsilon$). 
Note that, in this approximation, the difference between coframe and
coordinate indices  completely disappears.
This justifies our choice, in (\ref{la.2}) and in the sequel, of the same notation for these  (basically different) indices.

In accordance with (\ref{la.3}), only for weak coordinate transformations are considered.
Under a shift 
\begin{equation}\label{la.4}
x^a\mapsto x^a+\xi^a(x)\,,
\end{equation}
the components of the coframe are transformed  as
\begin{equation}\label{la.5}
{h^a}{}_b\mapsto {h^a}_{b}-{\xi^a}_{,b}\,.
\end{equation}
Thus,  in order to preserve the weakness of the fluctuation,  
it is necessary to require ${\xi^a}_{,b}=O(||{h^a}_{b}||)$. 
We will use the term   {\it   approximately covariant} \cite{ABS} 
for the expressions which are covariant only 
to the first order of the perturbations.
Observe that this assumption restricts only the amplitudes of
the perturbations and of their derivatives.
It does not restrict, however, the local freedom to transform the coordinates.
An  appropriative coordinate system can still be chosen in a small neighborhood
of the identity transformation in order  to simplify the (local) field equations.

Similarly, in order to be in agreement
with the approximation condition (\ref{la.3}),
the global $SO(1,3)$ transformations of the coframe field, 
$\vt^a\mapsto {A^a}_b\vt^b$, have  also to be restricted.
It is enough to require  the
transformations to be in a small neighborhood of the identity
 \begin{equation}\label{la.5xx}
{A^a}_b=\delta^a_b+\alpha^a_b\,,\qquad ||\alpha^a_b||=o(1)\,.
\end{equation}
\subsection{Reduction of the field}
In (\ref{la.2}), ${h^a}_b$ is a perturbation of the flat coframe. Thus
\begin{itemize}
\item[{(i)}]
To the first order, the holonomic coframe is expressed by the unholonomic one as
\begin{equation}\label{la.6}
dx^a=(\d^a_b-{h^a}_b)\vt^b\,.
\end{equation}
\item[{(ii)}]
The indices in ${h^a}_{b}$ can be lowered and raised 
by the Minkowskian metric 
\begin{equation}\label{la.7}
h_{ab}:=\eta_{am}{h^m}_{b}\,, \qquad h^{ab}:=\eta^{bm}{h^a}_{m}\,.
\end{equation}
The first operation is exact (covariant to all orders of approximations), while 
the second  is covariant only  to the first order, when $g^{ab}\approx \eta^{ab}$.
\item[{(iii)}]
The symmetric and the antisymmetric combinations of the  perturbations
\begin{equation}\label{la.9}
\theta_{ab}:=h_{(ab)}=\frac 12 (h_{ab}+h_{ba}),\qquad {\text{and}}\qquad
w_{ab}:=h_{[ab]}=\frac 12 (h_{ab}-h_{ba})\,.
\end{equation}
as well as the  trace
$\theta:={h^m}_m={\theta^m}_m$
are covariant to the first order.
\item[{(iv)}]
 The components of the metric tensor, in the linear approximation, involve only the
symmetric combination of the coframe perturbations
\begin{equation}\label{la.12}
g_{ab}=\eta_{ab}+2\theta_{ab}\,.
\end{equation}
\item[{(v)}]
Under the transformations (\ref{la.4}), two covariant pieces of the fluctuation change as
\begin{equation}\label{la.13}
\theta_{ab}\mapsto\theta_{ab}-\xi_{(a,b)}\,,\qquad {\text{and}}\qquad
w_{ab}\mapsto w_{ab}-\xi_{[a,b]}\,.
\end{equation}
 \end{itemize}
Thus the  approximately covariant irreducible
decomposition of the dynamical variable
\begin{equation}\label{la.13x}
h_{ab}=\theta_{ab}+w_{ab}\,.
\end{equation}
is obtained. 
Thus, instead of one field $h_{ab}$, we have, in this approximation, 
two independent fields: a symmetric field  
$\theta_{ab}$ and  an antisymmetric field $w_{ab}$.

\subsection{Gauge conditions}
The actual values of the components of the fields  $\theta_{ab}$ and $w_{ab}$
depend on a choice of a  coordinate system.
Thus four arbitrary relations between the components
(equal to the number of coordinates) may be imposed.
We require these relations to be Lorentz invariant, i.e., covariant in the 
first order approximation.
Thus the most general form of constraints (gauge conditions) that 
involve the first order derivatives is
 \begin{equation}\label{g.1}
\alpha \,{\theta_{am}}^{,m}+\beta \,\theta_{,a}+\gamma \,{w_{am}}^{,m}=0\,,
\end{equation}
where $\alpha,\beta,\gamma$ are dimensionless parameters.

Certainly, for some special values of the parameters, these conditions cannot
be realized.
Indeed, under the coordinate transformations (\ref{la.4}),  Eq. (\ref{g.1}) changes, in the lowest order, to
\begin{equation}\label{g.2}
\alpha \,\tilde{\theta}_{am}\,^{,m}+\beta \,\tilde{\theta}_{,a}+
\gamma \,\tilde{w}_{am}\,^{,m}=
\left(\alpha \,{\xi_{(a,m)}}+\beta \, \xi_{m,a}+\gamma \,{\xi_{[a,m]}}\right)^{,m}\,.
\end{equation}
Thus the conditions (\ref{g.1}) can be realized, by the coordinate transformations 
(\ref{la.4}),
if and only if  the system of PDE (\ref{g.2}) has a solution $\xi(x)$ for a given LHS.
 
Let us check the integrability of this system.
Eq. (\ref{g.2}) results in 
\begin{equation}\label{g.2x}
\left(\alpha \,{\xi_{(a,m),b}}+\beta \, \xi_{m,a,b}+\gamma \,{\xi_{[a,m],b}}\right)^{,m}
=\alpha \,\tilde{\theta}_{am,b}\,^{,m}+\beta \,\tilde{\theta}_{,a,b}+
\gamma \,\tilde{w}_{am,b}\,^{,m}\,.
\end{equation}
Commuting the  indices $a$ and $b$,  we obtain
\begin{equation}\label{g.3}
(\alpha+\gamma)\square \,\xi_{[a,b]}=2(\alpha\,\theta_{m[a,b]}-\gamma\, w_{m[a,b]})^{,m}\,.
\end{equation}
Thus, the gauge condition (\ref{g.1}) with
$\alpha=-\gamma\ne 0$
 cannot be realized by any change of the coordinate system.

\noindent Now, take the trace of (\ref{g.2x}) 
\begin{equation}\label{g.4}
(\alpha+\beta)\square\, {\xi_m}^{,m}=\alpha\,{\theta_{mn}}^{,m,n}+\beta \,\square \,\theta \,.
\end{equation}
Thus 
$\alpha=-\beta\ne 0$
is also forbidden. 

We will apply, in the sequel, two separate gauge conditions: for the symmetric field
 \begin{equation}\label{g.5}
{\theta_{am}}^{,m}-\frac 12 \theta_{,a}=0\,,
\end{equation}
and for the antisymmetric field
 \begin{equation}\label{g.6}
{w_{am}}^{,m}=0\,.
\end{equation}
Observe, that (\ref{g.5}) and   (\ref{g.6}) cannot be realized simultaneously 
by the same coordinate transformation.
Indeed, for this, the coordinate functions have to satisfy
 \begin{equation}
\square\, \xi_a= 2{\theta_{am}}^{,m}-\theta_{,a}\,,\\
\qquad {\textrm {and}} \qquad
\square\, \xi_a-({\xi_m}^{,m})_{,a}={w_{am}}^{,m}\,.
\end{equation}
The integrability conditions for these equations yield
\begin{equation}
\square\, \xi_{[a,b]}=2{\theta_{m[a,b]}}^{,m}=-{w_{m[a,b]}}^{,m}\,.
\end{equation}
For arbitrary independent fields $\theta_{ab}$ and $w_{ab}$, these conditions
are not satisfied. 

\noindent Certainly, the conditions (\ref{g.5}) and (\ref{g.6}) can be realized, separately, 
by  transformation of the coordinates. 
\subsection{Reduction of the field strengths}              
By (\ref{la.3}), let us decompose the field strengths  (\ref{tg.7} -- \ref{tg.9}).  
The 2-form $\C_a$ is approximated by 
\begin{equation}\label{fs.1}
\C_a=h_{ab,c}\,dx^c\wedge dx^b=-h_{a[b,c]}\,\vt^b\wedge \vt^c=
-(\theta_{a[b,c]}+w_{a[b,c]})\,\vt^b\wedge \vt^c\,.
\end{equation}
Consequently, the first part of the  field strength, (\ref{tg.7}), 
takes the form
\begin{equation}\label{fs.2}
{}^{(1)}\F_a=
-(\rho_1+\rho_3)(\theta_{a[b,c]}+w_{a[b,c]})\,\vt^b\wedge \vt^c\,.
\end{equation}
As for the second part, (\ref{tg.8}), it involves only the antisymmetric field,
\begin{equation}\label{fs.3}
{}^{(2)}\F_a=-3\rho_2w_{[ab,c]}\,\vt^b\wedge \vt^c\,.
\end{equation}
The third part,  (\ref{tg.9}), takes the form
\begin{equation}\label{fs.4}
{}^{(3)}\F_a=
\rho_3\eta_{ac}({h_{mb}}^{,m}-h_{,b})\,\vt^b\wedge \vt^c=
\rho_3\eta_{ac}({\theta_{bm}}^{,m}-\theta_{,b}-{w_{bm}}^{,m})\,
\vt^b\wedge \vt^c\,.
\end{equation}
Therefore, the  field strength is reduced to the sum of two
independent strengths --- one defined by  the symmetric field $\theta_{ab}$ 
and the second one defined by the antisymmetric field $w_{ab}$
\begin{equation}\label{fs.5}
\F_a(\theta_{mn},w_{mn})={}^{(\tt sym)}\F_a(\theta_{mn})+{}^{(\tt ant)}\F_a(w_{mn})\,,
\end{equation}
where
\begin{equation}\label{fs.5a}
{}^{(\tt sym)}\F_a=-\left[(\rho_1+\rho_3)\theta_{a[b,c]}+
\rho_3\eta_{a[b}{\theta_{c]m}}^{,m}-\rho_3\eta_{a[b}\theta_{,c]}\right]\vt^b\wedge \vt^c\,,
\end{equation}
and
\begin{equation}\label{fs.5b}
{}^{(\tt ant)}\F_a=-\left[(\rho_1+\rho_3)w_{a[b,c]}+3\rho_2w_{[ab,c]}-
\rho_3\eta_{a[b}{w_{c]m}}^{,m}\right]\vt^b\wedge \vt^c\,.
\end{equation}
Hence, for arbitrary values of the parameters $\rho_i$, the field
strengths are independent. 

\subsection{Reduction of the field equations}              
The field equation (\ref{tg.11}) includes the second order derivatives
of the perturbations in its LHS 
and the squares of the first order derivatives in  both sides.
In the linear approximation (\ref{la.3}), the quadratic terms can be neglected.
Thus,  (\ref{tg.11}) is approximated by  
\begin{equation}\label{sfe.1}
d*\F^a=0\,.
\end{equation}
The  covector valued 2-form $\F_a$  can be
expressed in the unholonomic basis as
$\F_a= F_{abc}\vt^b\wedge \vt^c/2$.
Accordingly, we derive
$$
d*\F^a= \frac 12 F_{abc,m}dx^m\wedge *(\vt^b\wedge \vt^c)=
- \frac 12 {F_{abc}}^{,m}*\big[e_m\rfloor (\vt^b\wedge \vt^c)\big]= \frac 12{F_{a[bc]}}^{,c}*\vt^b\,.
$$
Consequently, Eq. (\ref{sfe.1}) reads
\begin{equation}\label{sfe.2}
{F_{a[bc]}}^{,c}=0\,.
\end{equation}
Applying the antisymmetrization of the corresponding indices 
to the expression (\ref{fs.5}) we derive
the linearized field equation
\br \label{sfe.3}
&& (\rho_1+\rho_3)\big(\square\,\theta_{ab}-{\theta_{am,b}}^{,m}\big)+
\rho_3\big(-\eta_{ab}\square\,\theta
  -{{\theta_{mb}}^{,m}}_{,a}+\theta_{,a,b}+\eta_{ab}{\theta_{mn}}^{,m,n}\big)+\nonumber\\
&& \qquad  (\rho_1+2\rho_2+\rho_3) \big(\square \,w_{ab}-{w_{am,b}}^{,m}\big)+
(2\rho_2+\rho_3){w_{bm,a}}^{,m}=0\,.
\er
\pro\label{prop2}
For the case  $\rho_1=0$,  the linearized coframe field equation (\ref{sfe.3}),
in arbitrary coordinates, splits  into  two independent systems
$${}^{(\tt sym)}\E_{(ab)}(\theta_{mn})=0\,,
\qquad {\textrm {and}} \qquad
{}^{(\tt ant)}\E_{[ab]}(w_{mn})=0\,.$$
If  $\rho_1\ne 0$, Eq.(\ref{sfe.3}) does not split in
any coordinate system.
\epro
\prf
The equation (\ref{sfe.3}) is  tensorial to the first order.
Thus, by applying symmetrization and antisymmetrization operations, it
is reduced covariantly to a system of two independent
tensorial (to the first order) equations.
The symmetrization yields a system of 10 independent equations
\begin{equation}\label{sfe.4}
\square \,\big[(\rho_1+\rho_3)\theta_{ab}-\rho_3\eta_{ab}\theta\big]
- (\rho_1+2\rho_3){\theta_{m(a,b)}}^{,m}+
\rho_3(\theta_{,a,b}+\eta_{ab}{\theta_{mn}}^{,m,n})+
\rho_1{w_{m(a,b)}}^{,m}=0\,.
\end{equation}
The antisymmetrization yields a system of 6 independent equations
\begin{equation}\label{sfe.5}
(\rho_1+2\rho_2+\rho_3)\square\, w_{ab}
+(\rho_1+4\rho_2+2\rho_3){w_{m[a,b]}}^{,m}-\rho_1{\theta_{m[a,b]}}^{,m}=0\,.
\end{equation}
Evidently, the condition $\rho_1=0$ removes the ``mixed terms'' and yields the
separation of the system. Such splitting holds in arbitrary system of coordinates.

Suppose now $\rho_1\ne 0$. 
Thus, the "mixed terms" remain in both equations --- 
the $w$-term in (\ref{sfe.4}) and 
the $\theta$-term in (\ref{sfe.5}). 
Let us try to remove these terms by an appropriative choice of a coordinate system. 
For this we have to require the equations
$${\theta_{m[a,b]}}^{,m}=0, \qquad {\textrm {and}} \qquad
{w_{m(a,b)}}^{,m}=0\,$$
to hold simultaneously.
These equations can be satisfied only if
\begin{equation}\label{sfe.6}
{\theta_{ma}}^{,m}=0, \qquad {\textrm {and}} \qquad
{w_{ma}}^{,m}=0\,.
\end{equation}
The actual values of the variables $\theta_{ab}$ and $w_{ab}$
 depend on a choice of a  coordinate system.
Recall that the approximation conditions (\ref{la.3}) do not restrict
the freedom to choose  the local coordinate transformations.
Therefore, by (\ref{la.4}), four additional conditions (equal to the
number of coordinates), can still be applied to the perturbations in
order to satisfy (\ref{sfe.6}).
We need, however,  to eliminate eight independent expressions
${w_{ma}}^{,m}$ and ${\theta_{ma}}^{,m}$.
This cannot be done by four independent functions of the coordinates.
Indeed, under the transformations (\ref{la.4}), 
\br\label{sfe.7}
{\theta_{ma}}^{,m}&\mapsto&{\theta_{ma}}^{,m}-{\xi_{(m,a)}}^{,m}\,,\\
\label{sfe.8}
{w_{ma}}^{,m}&\mapsto&{w_{ma}}^{,m}-{\xi_{[m,a]}}^{,m}\,.
\er
Hence  the coordinate transformations have to satisfy
\begin{equation}\label{sfe.9}
{\xi_{(m,a)}}^{,m}={\theta_{ma}}^{,m}\,, \qquad {\textrm {and}} \qquad
{\xi_{[m,a]}}^{,m}={w_{ma}}^{,m}
\end{equation}
simultaneously.
Therefore,
\begin{equation}\label{sfe.9x}
{\xi_{m,a}}^{,m}={h_{ma}}^{,m}\,.
\end{equation}
The consistency condition for (\ref{sfe.9x}) is
$$
{h_{ma,b}}^{,m}={h_{mb,a}}^{,m}\,,
$$
which it is not satisfied in general.
\qed

Consequently, for $\rho_1=0$ and generic values of the parameters $\rho_2,\rho_3$,
the field equation of the coframe field is reduced
to two independent field equations for  independent field variables. 

(i) The symmetric field $\theta_{ab}$ of 10 independent variables satisfies the system of 10 independent equations 
\begin{equation}\label{sfe.10}
{}^{(\tt sym)}\E_{(ab)}(\theta_{mn}):=\rho_3\left[\square(\theta_{ab}-\eta_{ab}\theta)-{\theta_{m(a,b)}}^{,m}+
\theta_{,a,b}+\eta_{ab}{\theta_{mn}}^{,m,n}\right]=0\,.
\end{equation}
We  rewrite it as
\begin{equation}\label{sym.6}
\square\left(\theta_{ab}-\eta_{ab}\theta\right)
- \left({\theta_{am}}^{,m}-\frac 12 \theta_{,a}\right)_{,b}
-\left({\theta_{bm}}^{,m}-\frac 12 \theta_{,b}\right)_{,a}+
\eta_{ab}{\theta_{mn}}^{,m,n}=0\,.
\end{equation}
Substituting here the condition (\ref{g.5}) and its consequence
\begin{equation}\label{sym.7}
{\theta_{mn}}^{,m,n}=\frac 12 \square \ \!\theta\,
\end{equation}
we obtain 
\begin{equation}\label{sym.4}
\square \left(\th^{ab}- \frac 12\eta^{ab}\th\right)=0\,.
\end{equation}
Eq. (\ref{sym.4}) results in $\square \ \!\theta=0$. Then it is equivalent to 
\begin{equation}\label{sym.8}
\square\, \theta_{ab}=0\,.
\end{equation}
Consequently, in the coordinates associated with (\ref{g.5}), the  
symmetric field satisfied the wave equation. 

(ii) The antisymmetric system of 6 independent equation for 6 independent variables
\begin{equation}\label{sfe.11}
{}^{(\tt ant)}\E_{[ab]}(w_{mn}):=(2\rho_2+\rho_3)\left(\square\, w_{ab}+2{w_{m[a,b]}}^{,m}\right)=0\,.
\end{equation}
In the coordinates associated with (\ref{g.6}) it is reduced to the wave equation 
\begin{equation} \label{wr.11}
\square \, w_{ab}=0\,.
\end{equation}
\subsection{Reduction of the Lagrangian}
In the sequel of this paper, we consider the models with parameter $\rho_1=0$.
Let us examine now  the  reduction of the  Lagrangian (\ref{tg.5}).
\pro
For $\rho_1=0$, the Lagrangian of the coframe field
is reduced, up to a total derivative term, to the sum of two independent Lagrangians
\begin{equation}\label{rl.2}
\L(\theta_{ab},w_{ab})={}^{\tt (sym)}\L(\theta_{ab})+{}^{\tt (ant)}\L(w_{ab})\,.
\end{equation}
\epro
\prf
With $\rho_1=0$ the term  ${}^{(1)}\L$ does not appears in the Lagrangian.
Calculate in the linear approximation (
we use the abbreviation $\vt^{a b\cdots}=\vt^a\wedge\vt^b\wedge\cdots$) 
\begin{equation}\label{rl.3}
{}^{(2)}\L= (d\vt^a\wedge\vt_a)\wedge*(d\vt_b\wedge\vt^b)=
 h^{am,n}h_{bp,q}\vt_{nma}\wedge*\vt^{qpb}\,.
\end{equation}
Applying the formula
\begin{equation}\label{rl.4}
\vt_{abc}\wedge*\vt^{a'b'c'}=6\d_{a}^{[a'}\d_{b}^{b'}\d_{c}^{c']}*1\,
\end{equation}
we derive
\begin{equation}\label{rl.5}
{}^{(2)}\L=2w^{ab,c}(w_{ab,c}+w_{ca,b}+w_{bc,a})*1\,.
\end{equation}
So ${}^{(2)}\L$ depends only on the antisymmetric field.
Consider now the linear approximation to the term ${}^{(3)}\L$
\begin{equation}\label{rl.6}
{}^{(3)}\L=(d\vt_a\wedge \vt_b)\wedge *(d\vt^b\wedge\vt^a)=
 {h_a}^{m,n}{h^b}_{p,q}\vt_{nmb}\wedge*\vt^{qpa}\,.
\end{equation}
Use  (\ref{rl.4}) to get
\begin{equation}\label{rl.7}
{}^{(3)}\L= \left[h_{ab,c}(h^{ab,c}-{h^{ac,b}})-{h_{ab}}^{,a}{h^{cb}}_{,c}
+\theta^{,a}(2{h_{ba}}^{,b}-\theta_{,a})\right]*1\,.
\end{equation}
Insert here the splitting (\ref{la.13x}). 
It follows that the Lagrangian (\ref{rl.7}) is reduced to the sum
\begin{equation}\label{rl.8}
{}^{(3)}\L={}^{(3)}\L(\theta)+{}^{(3)}\L(w)+{}^{(3)}\L(\theta,w),
\end{equation}
where
\br\label{rl.9}
{}^{(3)}\L(\theta)&=& \left[\theta_{ab,c}(\theta^{ab,c}-{\theta^{ac,b}})-
{\theta_{ab}}^{,a}{\theta^{cb}}_{,c}+\theta^{,a}(2{\theta_{ba}}^{,b}-\theta_{,a})\right]*1\,,\\
\label{rl.10}
{}^{(3)}\L(w)&=&
 \left[w_{ab,c}(w^{ab,c}-{w^{ac,b}})-{w_{ab}}^{,a}{w^{cb}}_{,c}\right]*1\,,\\
\label{rl.11}
{}^{(3)}\L(\theta,w)&=&2\left[-\theta_{ab,c}{w^{ac,b}}+\theta^{,a}{w_{ba}}^{,b}-
{\theta_{ab}}^{,a}{w^{cb}}_{,c}\right]*1\,.
\er
Extracting  the total derivatives in the mixed term (\ref{rl.11}) we obtain
\begin{equation}\label{rl.12}
{}^{(3)}\L(\theta,w)=\left(\theta_{ab}({w^{ac,b}}-{w^{bc,a}})_{,c}
-\theta{w_{ba}}^{,a,b}\right)*1+\, {\textrm{exact terms}}\,.
\end{equation}
The terms in the brackets vanish identically as a product of symmetric and
antisymmetric tensors. Thus the mixed term ${}^{(3)}\L(\theta,w)$ is a total derivative. 
Consequently,  desired reduction of the Lagrangian is obtained.
\qed

The Lagrangian of the symmetric field ${}^{\tt (sym)}\L={}^{(3)}\L(\theta)$
may be rewritten in a more compact form.
Observing the identity
\begin{equation}\label{rl.13}
{\theta_{ab}}^{,a}{\theta^{cb}}_{,c}=
{\theta_{ab}}^{,c}{\theta^{cb}}_{,a}+\, {\textrm{exact terms}}\,,
\end{equation}
and extracting the total derivatives, we obtain
\begin{equation}\label{rl.14}
{}^{\tt (sym)}\L=\frac 12 \rho_3\left[\theta_{ab,c}(\theta^{ab,c}-2{\theta^{ac,b}})
+\theta^{,a}(2{\theta_{ba}}^{,b}-\theta_{,a})\right]*1\,.
\end{equation}
This form of the Lagrangian is acceptable in arbitrary coordinates. 
In the coordinates associated with the condition (\ref{g.5}), the last brackets in (\ref{rl.14})
vanish.
In the first brackets, we extract the total derivatives and use 
 (\ref{g.5}) to derive
(symbol $\approx$ used here for equality up to total derivatives)
$$
\theta_{ab,c}{\theta^{ac,b}}=(\theta_{ab}{\theta^{ac,b}})_{,c}-
\theta_{ab}{{\theta^{ac,b}}}_{,c}\approx -\frac 12\theta_{ab}\theta^{,a,b}
\approx \frac 12{\theta_{ab}}^{,b}\theta^{,a}\approx \frac 14{\theta_{,a}}\theta^{,a}\,.
$$
Consequently the symmetric field Lagrangian (\ref{rl.9}) is reduced to
\begin{equation}\label{sym.3}
{}^{({\rm sym})}\L=\frac 12 \kappa\Big(\th_{ab,c}\th^{ab,c}-
\frac 12 \th_{,a}\th^{,a}\Big)*1\,.
\end{equation}

Analogously, for the Lagrangian of the antisymmetric field
${}^{\tt (ant)}\L={}^{(2)}\L+{}^{(3)}\L(w)$, we use the identity
\begin{equation}\label{rl.15}
{w_{ab}}^{,a}{w^{cb}}_{,c}={w_{ab,c}}{w^{ac,b}}+\, {\textrm{exact terms}}\,
\end{equation}
and rewrite it, in an arbitrary system of coordinates, as
\begin{equation}\label{rl.16}
{}^{\tt (ant)}\L=\frac 12 (2\rho_2+\rho_3) \left[w_{ab,c}(w^{ab,c}-2{w^{ac,b}})\right]*1\,,
\end{equation}
or, equivalently, as 
$$
\L(w)=\frac 12 (2\rho_2+\rho_3)\Big(w_{ab,c}(w^{ab,c}-{w^{ac,b}})-
{w_{ab}}^{,a}{w^{cb}}_{,c}\Big)*1\,.
$$
The gauge condition (\ref{g.6}) removes the last term while 
the second term is rewritten as
$$w_{ab,c}{w^{ac,b}}\approx -w_{ab}{w^{ac,b}}_{,c} \approx 0\,.$$
Thus, the Lagrangian of the antisymmetric field is
\begin{equation}\label{rl.14x}
{\tilde{\L}}(w)=\frac 12(2\rho_2+\rho_3)w_{ab,c}w^{ab,c}*1\,.
\end{equation}
\subsection{Reduction  of the energy-momentum current}
The Lagrangian of the coframe field is decomposed, in the first order approximation,
to a sum of two independent Lagrangians for two independent fields.
The Noether current expression, being derivable from the Lagrangian,
has to have the same splitting.
\pro
The coframe energy-momentum current  is reduced, on shell, in the first order approximation, as
\begin{equation}\label{em.1}
\T_a(\theta_{mn},w_{mn})= {}^{\tt (sym)}\T_a(\theta_{mn})+{}^{\tt (ant)}\T_a(w_{mn})\,,
\end{equation}
up to a total derivative.
\epro
\prf
The coframe energy-momentum current is of the form
\begin{equation}\label{em.2}
\T_a=(e_a\rfloor \C_m)\wedge *\F^m -e_a\rfloor \L\,.
\end{equation}
Due to Proposition 2, the second term,  in the first order  approximation,
does not contain the mixed terms $\theta' \cdot w'$.
Hence, it already has the reduced form.
To treat the first term,
we write  the strengths in the component
\begin{equation}\label{em.3}
\C_m= C_{m[bc]}\vt^b\wedge\vt^c, \qquad \F^m= {F^m}_{[pq]}\vt^p\wedge\vt^q\,.
\end{equation}
Thus, the first term of (\ref{em.2}) is approximated by
\begin{equation}\label{em.4}
(e_a\rfloor \C_m)\wedge *\F^m=
C_{m[bc]}{F^m}_{[pq]}(e_a\rfloor\vt^{bc})\wedge*\vt^{pq}
=4C_{m[an]}F^{m[bn]}*\vt_b=4h_{m[a,n]}F^{m[bn]}*\vt_b\,.
\end{equation}
The  3-form $*\vt_b$, in the lowest order approximation, is an exact form.
Thus, it is enough to show that the 
scalar factor, in the RHS of (\ref{em.4}), has the desired splitting. 
This expression is a sum of two terms. The first one is proportional to
$$h_{ma,n}F^{m[bn]}= - h_{ma}{F^{m[bn]}}_{,n}+{\textrm {total derivatives\,,}}$$
i.e., it is, on shell, an exact form.
Now we have to show that the second term, which is proportional to
$h_{ma,n}F^{m[bn]}$,
does not involve the mixed products of a type $\theta\cdot w$.
The mixed product expression in the latter term is proportional to
\begin{equation}\label{em.4x}
\theta_{mn,a}(w^{mb,n}+2\eta^{m[n}{w^{b]k}}_{,k})
+w_{mn,a}(\theta^{mb,n}+\eta^{mb}{\theta^{nk}}_{,k}-\eta^{mb}\theta^{,n})\,.
\end{equation}
By recollection of the terms, we rewrite this expression as
\begin{equation}\label{em.4xx}
(\theta_{mn,a}w^{mb,n}+{\theta_{mn}}^{,n}{w^{bm}}_{,a})+
(\theta_{,a}{w^{bm}}_{,m}-\theta_{,m}{w^{bm}}_{,a})+
(\theta^{mb,n}w_{mn,a}-{\theta^{bm}}_{,a}{w_{mn}}^{,n})\,.
\end{equation}
The three brackets above are total derivatives, namely,
\begin{equation}\label{em.4xxx}
\left[(\theta_{mn,a}w^{mb})^{,n}+({\theta_{mn}}^{,n}{w^{bm}})_{,a}\right]
+\left[(\theta{w^{bm}}_{,m})_{,a}-(\theta{w^{bm}}_{,a})_{,m}\right]+
\left[(\theta^{mb,n}w_{mn})_{,a}-({\theta^{bm}}_{,a}{w_{mn}})^{,n}\right]\,.
\end{equation}
Thus, (\ref{em.4}) and, consequently, (\ref{em.2}) do not involve the mixed terms. 
The desired splitting is proved.
\qed 

The energy-momentum tensor ${T_{a}}^b$ can be derived from the Noether current 
$\T_a$ by applying the relations 
\begin{equation}\label{cur-ten}
\T_a={T_{a}}^b*\vt_b\,,\qquad T_{ab}=e_b\rfloor *\T_a\,.
\end{equation} 
\pro
For the field $\theta_{ab}$ in the coordinate system associated with the gauge condition
\begin{equation}\label{em.6}
{\theta_{am}}^{,m}-\frac 12 \theta_{,a}=0\,,
\end{equation}
the  energy-momentum tensor  is 
\begin{equation}\label{sym.12}
T_{ab}=\frac 12 \kappa\left[\left(\th_{mn,a}{\th^{\,mn}}_{,b}-
\frac 14 \eta_{ab}\th_{lm,n}\th^{\,lm,n}\right)-\frac 12\left(\th_{,a}\th_{,b}-
\frac 14 \eta_{ab}\th_{,m}\th^{\,,m}\right)\right]\,.
\end{equation}
This tensor is symmetric and traceless. 
\epro
\prf
We start with the energy-momentum current for the coframe field 
$$\T_a=(e_a\rfloor \C_m)\wedge *\F^m -e_a\rfloor \L\,.$$
Due to Proposition 3, in the first order approximation, 
this current is decomposed to two independent 
currents.  
Thus we may assume $w_{ab}=0$ in order to derive the expression for $\T_a(\theta)$.

In the coordinates associated with the gauge condition (\ref{em.6}), 
by (\ref{sym.3}) 
$$
e_a\rfloor \L=\frac 12 \rho_3\left(\theta_{mn,p}\theta^{mn,p}
-\frac 12 \theta_{,m}\theta^{,m}\right)*\vt_a\,.
$$
The first term of $\T_a$ is derived from (\ref{em.2})
$$(e_a\rfloor \C_m)\wedge *\F^m=4\theta_{m[a,n]}F^{m[bn]}*\vt_b=
2\left(\theta_{ma,n}F^{m[bn]}*\vt_b-\theta_{mn,a}F^{m[bn]}*\vt_b\right)\,.
$$
Observe that, on shell, up to a total derivative
$$\theta_{ma,n}F^{m[bn]}\approx -\theta_{ma}{F^{m[bn]}}_{,n}=0\,.$$
Thus,
$$(e_a\rfloor \C_m)\wedge *\F^m=-2\theta_{mn,a}F^{m[bn]}*\vt_b\,.$$
Applying the gauge condition to (\ref{fs.5}) we get
$$\F_a=-\rho_3\left[\theta_{a[b,c]}+
\eta_{a[b}({\theta_{c]m}}^{,m}-\theta_{,c]})\right]\vt^{bc}=
-\rho_3\Big(\theta_{a[b,c]}-\frac 12 \eta_{a[b}\theta_{,c]}\Big)\vt^{bc}\,.
$$
Consequently,
$$(e_a\rfloor \C_m)\wedge *\F^m=
2\rho_3\theta_{mn,a}\Big(\theta^{m[b,n]}-\frac 12 \eta^{m[b}\theta^{,n]}\Big)*\vt_b$$
Extracting the total derivatives
\brn
&&\theta_{mn,a}\theta^{mb,n}\approx{\theta_{mn}}^{,n}{\theta^{mb}}_{,a}
\approx \frac 12 \theta_{,m}{\theta^{mb}}_{,a}\approx \frac 14 \theta_{,a}\theta^{,b}\,,\\
&&\theta_{mn,a}\eta^{mb}\theta^{,n}\approx{\theta_{mn}}^{,n}\theta_{,a}\approx
\frac 14 \theta_{,a}\theta^{,b}\,.
\ern
it follows that 
$$(e_a\rfloor \C_m)\wedge *\F^m=
\rho_3\Big(-2\theta_{mn,a}\theta^{mn,b}+\theta_{,a}\theta^{,b}\Big)*\vt_b$$
Collecting the terms into $\T_a$ and extracting the energy-momentum tensor
${T_{a}}^b$ from the current $\T_a$ by
$T_{ab}=e_b\rfloor *\T_a$
we get the desired expression. It is clear that energy-momentum 
tensor is symmetric and traceless. 
\qed

In GR, the behavior of small perturbations of the metric tensor is managed 
by the wave equation. 
 Thus, for a wave propagating   in the positive direction of the $x$-axis, 
only two independent components of the matrix $\th_{ab}$ remain. 
\begin{equation}\label{5.1}
\theta_{23}=\mu (\tau)\,, \qquad \theta_{22}=-\theta_{33}=\nu(\tau)\,, 
\qquad {\rm where} \quad
\tau=t-x\,.
\end{equation}
The calculation of the energy-momentum tensor for the symmetric field by use of the 
tensor (\ref{sym.12}) yields 
\begin{equation}\label{5.2}
T_{ab}=k(\mu_{,a}\mu_{,b}+\nu_{,a}\nu_{,b})\,.
\end{equation}
The energy flux reads 
\begin{equation}\label{5.3}
T_{01}=-\rho_3\left(\dot{\theta}^2_{23}+\frac 14 (\dot{\theta}_{22}-
\dot{\theta}_{33})^2\right)
\end{equation}
Observe that the expressions (\ref{5.2},\ref{5.3}) are the same 
as the expressions obtained in GR from  the energy-momentum pseudorensors. 

Let us turn now to the antisymmetric field.  
\pro
In the coordinate system associated with the gauge condition
\begin{equation}\label{em.8}
{w_{am}}^{,m}=0\,,
\end{equation}
the  energy-momentum tensor of the antisymmetric field is
\begin{equation}\label{em.9}
T_{ab}=-(2\rho_2+\rho_3)\left(w_{mn,a}{w^{mn}}_{,b}-\frac 14 \eta_{ab} w_{mn,p}w^{mn,p}\right)\,.
\end{equation}
This tensor is traceless and symmetric.
\epro
\prf
The current of the symmetric and of the antisymmetric fields are decoupled. Thus 
we may assume $\theta_{ab}=0$.
In the coordinates associated with the gauge condition (\ref{em.8}), 
$$
e_a\rfloor \L=\frac 12(2\rho_2+\rho_3)w_{ab,c}w^{ab,c}*\vt_b\,.
$$
As for the first term of $\T_a(w)$ we derive from (\ref{em.4})
$$(e_a\rfloor \C_m)\wedge *\F^m=4w_{m[a,n]}F^{m[bn]}*\vt_b=
2(w_{ma,n}F^{m[bn]}-w_{mn,a}F^{m[bn]})*\vt_b\,.
$$
The first term vanishes, on shell, up to a total derivative, 
$$w_{ma,n}F^{m[bn]}\approx -w_{ma}{F^{m[bn]}}_{,n}=0\,.$$
Thus,
$$(e_a\rfloor \C_m)\wedge *\F^m=-2w_{mn,a}F^{m[bn]}*\vt_b\,.$$
Inserting the gauge condition  (\ref{em.8}) into (\ref{fs.5}) we derive
$$\F_a=-(\rho_3w_{a[b,c]}+3\rho_2w_{[ab,c]})\vt^{bc}\,.$$
Hence,
$$(e_a\rfloor \C_m)\wedge *\F^m=2(\rho_3w_{ma,n}w^{m[b,n]}+
3\rho_2w_{ma,n}w^{[mb,n]})*\vt_b\,.$$
Extract the total derivatives and use the gauge condition to get  
\brn
&&w_{mn,a}w^{mb,n}\approx{w_{mn}}^{,n}{w^{mb}}_{,a}\approx 0\\
&&w_{mn,a}w^{bn,m}\approx{w_{mn}}^{,m}{w^{bn}}_{,a}\approx 0 \,.
\ern
Consequently, 
$$
(e_a\rfloor \C_m)\wedge *\F^m=-(2\rho_2+\rho_3)w_{mn,a}w^{mn,b}
$$
The desired expression (\ref{em.9})  is obtained now by collecting the terms. 
\qed 
\section{The role of the parameters $\rho_i$} 
The case $\rho_1=0$ is extracted in coframe models by existence of a unique spherical symmetric static solution.  Since the exact solution yields the Schwarzschild metric this condition generates a viable subclass of gravity coframe models.

We have involved an independent criteria. Namely, we have shown that only in  the case $\rho_1=0$ the weak perturbations of the coframe reduce to two independent fields with their own Lagrangian dynamics. 
Consequently the models have a free field limit. 
This effect is correlated to the resent obtained result \cite{Sousa:2003fv} concerning 
the Hamiltonian dynamics behavior. 

It is interesting to note that in $2D$ coframe gravity  only one term in the Lagrangian preceded by $\rho_1$ appears. Thus the corresponded reduction of fields is impossible. 


\end{document}